\documentclass[aps,twocolumn,floatfix]{revtex4}
\usepackage{graphicx}
\usepackage{times}
\newcommand{\vect}[1]{\mbox{\boldmath $#1$}}
\newcommand{\D}{{\rm d}}
\newcommand{\E}{{\rm e}}
\newcommand{\I}{{\rm i}}

\begin{document}
\title{
Chemo-capillary instabilities of a contact line 
     }
\author{L.\ M.\ Pismen}
\affiliation{Department of Chemical Engineering and 
Minerva Center for Nonlinear Physics of
Complex Systems, Technion -- Israel Institute of Technology, 
Haifa 32000, Israel}

\begin{abstract}
Equilibrium and motion of a contact line are viewed as analogs of phase equilibrium and motion of an  interphase boundary. This point of view makes evident the tendency to minimization of the length of the contact line at equilibrium. The concept of line tension is, however, of limited applicability, in view of a qualitatively different relaxation response of the contact line, compared to a two-dimensional curve.  
Both the analogy and qualitative distinction extend to a non-equilibrium situation arising due to coupling with reversible substrate modification. Under these conditions, the contact line may suffer a variety of chemo-capillary instabilities (fingering, traveling and oscillatory), similar to those of dissipative structures in nonlinear non-equilibrium systems. The preference order of the various instabilities changes, however, significantly due to a different way the interfacial curvature is relaxed.  
\end{abstract}

\maketitle 

\section{Introduction}

Instabilities of a contact line are commonplace in various settings (for a recent review, see Refs.~\cite{bonn,matarev}). In many practical applications, instabilities are detrimental when they, for example, impair smooth coating or spotless dewetting. Instabilities may lead, however, to formation of fine patterns or enhanced spreading in microfluidic applications, and cause such fascinating phenomena as spontaneous motion in non-equilibrium systems. Most thoroughly studied instabilities are of hydrodynamic origin, being caused, for example, by enhanced gravitational or Marangoni driving at a perturbed contact line, and are most commonly related to formation of capillary ridges \cite{matarev}. Quantitative characterization of these instabilities is commonly based on lubrication approximation, and is strongly affected by the notorious contact line singularity resolved on a microscopic scale. 

We will concentrate here on a different kind of instability, caused by substrate modification rather than hydrodynamics, and therefore apt to occur at slow velocities and small scales. We start in Sect.~\ref{S1} we discussing an analogy between equilibrium of a fluid body bounded by the contact line and thermodynamic phase equilibrium in two dimensions, which is made evident by variational representation of lubrication equations accounting for the action of surface tension and disjoining pressure. We will argue, however, that this analogy does not justify extending the notion of two-dimensional line tension to the contact line, in view of a qualitatively different quasi-elastic response of the contact line dependent on three-dimensional effects hidden in the lubrication description. 

In the main part of the paper, we will develop the thermodynamic analogy in a different direction by considering chemo-capillary instabilities in a non-equilibrium setting arising due to coupling with reversible substrate modification. As a representative example, we will consider a set-up used in observations of spontaneous motion of droplets induced by surfactant adsorption \cite{j05l}. This system has been described by a model combining hydrodynamics in lubrication approximation with a linear reaction-diffusion equation for an adsorbed species, and studied both numerically \cite{th05} and analytically \cite{p06}. We reconsider it here as an analog of nonlinear models responsible for the formation of non-equilibrium structures \cite{book}. There are remarkable similarities here, leading to the same variety of instabilities, which include, in addition to traveling instability, fingering instability and oscillations commonly found in other physical settings. The character of transitions is, however, strongly influenced by the specific wavenumber dependence of the contact line relaxation response, different from the elastic response of an interphase boundary in two dimensions.       
 
We will derive equations of motion of a curved contact line in Sect.~\ref{S11} by combining the effective equation of motion of a rectilinear contact line derived through resolving the contact line singularity \cite{pe08} with the curvature response derived here in a local approximation and fitting the established theory \cite{bonn,deGennes85}. After the diffusive surfactant model is introduced in Sect.~\ref{S12}, the various instabilities of a contact line delimiting a static or moving semi-infinite fluid layer or bounding a circular droplet are analyzed in the following Sections.

\section{Equilibrium} \label{S1}

The evolution equation of the layer thickness $h$ in lubrication approximation can be written in a variational form \cite{p02}
\begin{equation}
\partial_t h = - \nabla \cdot \kappa(h) \nabla\, \frac{\delta W}{\delta h}, 
\label{jst2} \end{equation}  
The driving potential $W$ is of the Cahn--Hilliard type, appropriate for the case when the ``order parameter'' is conserved: 
\begin{equation}
W = \int \left[\frac{\gamma}{2} |\nabla h|^2 
 + \Phi(h)  \right]\D^2 \vect{x}.
\label{jst3}  \end{equation}

The two terms in the integrand are the interfacial energy with the surface tension $\gamma$ and the net adhesion energy $\Phi(h)$; $\nabla$ is the two-dimensional gradient operator. Other extrinsic interaction terms, such as gravity, having the same algebraic form, can be added here, but will be further assumed to be negligible, as they operate on far larger scales. The mobility coefficient $\kappa(h)$ depends on $h$ in a strongly nonlinear fashion when it is of hydrodynamic origin, e.g. 
$\kappa(h) = \mbox{$\frac{1}{3}$} \eta^{-1}h^3$ 
for the Stokes flow with the dynamic viscosity $\eta$ and no slip on the solid substrate. The validity of the lubrication hydrodynamic approximation used in derivation of Eq.~(\ref{jst2}), as well as of the concept of shape-independent adhesion energy, is formally restricted to films with a large aspect ratio, implying small contact angles. It is universally used nevertheless for the analysis of both equilibrium and non-equilibrium films leading to qualiatively correct results even outside its formal applicability limits.    

It is clear from the variational form of Eq.~(\ref{jst2}) that the functional $W$ is minimized in a stationary configuration, which is defined by the Euler--Lagrange equation
\begin{equation}
\gamma\nabla^2 h - \Pi(h)  = 0, 
\label{jst4}  \end{equation}
where  $\Pi(h) = \Phi'(h)$ is the disjoining pressure. One can consider two basic configurations. A \emph{macroscopic} system (or a part thereof) can be  viewed as an infinite layer asymptotically flat at spatial infinity. Practically, the layer may be shaped at distances far exceeding the range of adhesion forces by gravity or other macroscopic forces. The enormous macroscopic to microscopic scale ratio, which may reach 7--8 orders of magnitude, leaves ample space for an asymptotically flat region at intermediate distances. The asymptotic slope is identified then with the equilibrium contact angle.

Consider, for example, a commonly used expression for disjoining pressure of the type  
\begin{equation}  
\label{af}  
\Pi(h) = \frac{A}{h^3}\left[1-\left(\frac{h_f}{h}\right)^{n}\right], 
\end{equation}  
characterizing a partially wetting liquid with the net van der Waals interaction parameter $A$ forming a precursor film of the thickness $h_f$. The equilibrium contact angle $\theta_e$ is computed by solving the one-dimensional version of Eq.~(\ref{jst4}), 
\begin{equation}
 \gamma h''(x) - \Pi(h) = 0, 
 \quad h(-\infty)=h_f,  \quad h'(\infty)=\theta_e, 
\label{ppst}  \end{equation}
where $x$ is the coordinate normal to the contact line. Using the phase plane representation $h'(x)=p(h)$ and integrating yields 
\begin{equation}
\theta_e = \left[2 \Phi(h_f)/\gamma \right]^{1/2}. 
\label{th}  \end{equation}

A representative example of a disjoining pressure of a more complex form \cite{sharma89} combines wetting van der Waals and dewetting polar interactions:
\begin{equation}  
\label{shf}  
\Pi(h) = - \frac{A}{h^3} + B e^{-h/\ell}. 
\end{equation} 
In a dimensionless form with $h$ scaled by the attenuation scale $\ell$ and $\Pi$ scaled by $A/\ell^3$, this function depends on a single parameter $\chi=B\ell^3/A$:
\begin{equation}  
\label{sf}  
\Pi(h) = - h^{-3} + \chi e^{-h}. 
\end{equation}  
The condition $f(h_f)=0$, which has positive roots at $\chi> \chi_f =(e/3)^3 \approx 0.7439$, determines the thickness $h_f$ of a thin film in equilibrium with a flat macroscopic liquid layer. The equilibrium contact angle computed  
in the same way as (\ref{th}) is positive at $\chi > \chi_0 = e^2/8 \approx 0.9236$. 

An alternative configuration is a \emph{microdroplet} of a limited volume (or several microdroplets). The volume constraint can be treated by extracting from  Eq.~(\ref{jst3}) the term $\mu V = \mu \int h(\vect{x})\, \D^2 \vect{x}$, where the Lagrange multiplier $\mu$ plays a role of chemical potential. For the function (\ref{af}), the simplest solution of the respective Euler--Lagrange equation,
\begin{equation}
\gamma\nabla^2 h - \Pi(h) +\mu = 0, 
\label{jstmu}  \end{equation}
is a parabolic cap of a radius $R$ on the top of a precursor layer with the thickness somewhat exceeding $h_f$. The Lagrange multiplier $\mu$ can be identified then with the chemical potential or the vapor phase in equilibrium with the droplet of curvature radius $R/\theta_e$, i.e. with the vapor concentration just below the dew point when $R \gg h_f\theta_e$. For the function (\ref{sf}), the solutions at at $\chi < \chi_0$ and a suitable value of $\mu>0$ can take the form of a straight-line front separating domains with alternative values of the stationary film thickness or of a circular ``pancake'' or ``hole''. 
  
Both macroscopic and microscopic configurations must be stable when the length of the contact line is minimal. Equation (\ref{ppst}) or (\ref{jstmu}) with a suitable function $\Pi(h)$ can be recognized as equations of phase equilibrium with the layer thickness $h$ playing the role of the order parameter. Respectively, Eq.~(\ref{jst2}), is recognized as the Cahn--Hilliard equation governing relaxation to the equilibrium state. The only unconventional features are a strongly nonlinear dependence of the mobility on the order parameter and a possibility for the latter to be infinite in a system of infinite extent. This analogy allows us to consider the question of contact line stability in the general framework of stability of phase equilibria. 

The question of stability of different configurations of an equilibrium contact line has been discussed in connection with the notion of \emph{line tension} \cite{Dietrich07}. Unlike a line in two dimensions, deformations of a contact line inadvertently cause deformations of the surface of the enclosed fluid volume, and therefore instability may not arise even when the line tension is formally negative \cite{Mechkov07}. Moreover, line tension of a contact line, unlike surface tension in three or line tension in two dimensions, is not a fundamental property. If it is defined as a constituent part of the overall energy proportional to the line length, it remains dependent on the shape of the interface in the vicinity of the contact line, and therefore cannot be separated from surface tension and wetting properties. As noted in Ref.~\cite{Mechkov07}, stability of an equilibrium contact line is fully determined by the applicable dependence of the disjoining pressure on the local layer thickness $h$, which determines both the equilibrium contact angle and the interface shape in the vicinity of the contact line affecting the apparent line tension.  
Mechkov \emph{et al} \cite{Mechkov07} proved stability of a circular contact line and stability of a capillary ridge on short wavelengths. In the lubrication approximation, this result is evident: both the transitional area near the contact line and a front separating domains with different film thickness carry positive energy, and their length should be minimized by the globally stable configuration. 


\section{Equation of motion for the contact line} \label{S11}

Instabilities of a contact line or a front between alternative equilibrium thickness levels may arise in a one-component fluid only in a non-equilibrium setting. The key ingredient of the stability analysis in different situations is the equation of motion for the contact line. A naive approach based on the local curvature of the contact line and line tension that would be sufficient if the problem was truly two-dimensional is not applicable here, since the three-dimensional configuration of the interface in the vicinity of the contact line plays a crucial role. Strictly speaking, the problem is nonlocal, since any distortion of the contact line causes pressure changes within the liquid layer or droplet that affect all other locations along the contact line. The problem becomes tractable only when pressure fluctuations are neglected and the analysis is based on a local equation of contact line motion. Such an equation can be extracted from the analysis of slow displacement or spreading of droplets based on the solvability condition for perturbed stationary configurations \cite{pp04,p06,pe08}. The theory is based on matching of perturbations induced by displacement of the contact line in the macroscopic domain and in the contact line vicinity where the dynamic contact line singularity is resolved through either slip or the presence of a precursor film. For a droplet spreading due to a difference between the apparent macroscopic angle $\theta$ and the equilibrium contact angle $\theta_e$, the spreading velocity $U$ is \cite{pe08}
\begin{equation}
U = \frac{1}{3} Q U_0 (\theta^3 - \theta_e^3), \qquad  Q =  \frac{1}{3 \ln (aR/\ell)},
\label{uth}  \end{equation}
where $U_0=\gamma/\eta$ is the characteristic viscous velocity and the argument of the logarithm contains the ratio of a macroscopic length $R$ (droplet radius) to a microscopic length $\ell$ (thickness of the precursor layer or slip length); the numerical factor $a$ is dependent on the functional form of the disjoining pressure and/or other microscopic and macroscopic inputs. The linearized form of Eq.~(\ref{uth}) applicable when the difference $\vartheta=\theta - \theta_e$ is small is   
\begin{equation}
U =  Q  \theta_e^2 U_0 \vartheta.
\label{uth0}  \end{equation}
If this expression is applied to compute the displacement velocity of a droplet under the action of a variable equilibrium contact angle, the result coincides precisely with that obtained earlier in a somewhat different way \cite{p06}. We will further adopt Eq.~(\ref{uth}) and its linearization as the local equation of motion of the contact line also for non-circular geometry, neglecting nonlocal effects which are likely to manifest themselves only as weak corrections of the numerical geometric factor in the argument of the logarithm.

Perturbations of the apparent contact angle $\vartheta$ can be computed in the following way. Let the contact line be shifted from an unperturbed position $\Gamma(y)$ along the outer normal by an increment $\xi(y)$, where $y$ is a spanwise coordinate. The displacement is supposed to be small on the macroscopic scale, although it may be large compared to the microscopic scale $\ell$; the essential requirement is that the spanwise derivative $\xi'(y)$ be small, so that the corrugation wavelength be much larger than $\ell$. Projecting the perturbation upon $\Gamma$ implies the perturbation of the local film thickness  $\widetilde h(\Gamma,y) =  \theta_e \xi(y)$. The global perturbation is obtained by solving the two-dimensional Laplace equation $\nabla^2 \widetilde{h}=0$ in the \emph{unperturbed} geometry, and $\vartheta$ is subsequently computed as the normal derivative $\vect{n} \cdot \nabla \widetilde{h}=0$ at the contact line.     

Consider first relaxation of a perturbed rectilinear contact line bounding the liquid layer at $x<0$ with a constant incline $\theta_e$. The contact line perturbation is presented as a Fourier integral $\xi(y)= \int \zeta_k \cos ky\, dk$. For small deviations of the contact angle, the perturbation of the film thickness at the axis $x=0$ (i.e. at the nominal position of the undisturbed contact line) induced by a Fourier component with $k \ll\ell^{-1}$ is $\widetilde h(0,y)=  \theta_e \zeta_k \cos ky$. The perturbation of the film thickness  $\widetilde h(x,y)$ in the bulk of the film obtained by solving the Laplace equation with this boundary condition is 
\begin{equation}
 \widetilde h(x,y) =  \theta_e \zeta_k\, \E^{|k|x} \cos ky .
\label{hthx}  \end{equation}
This yields the perturbation of the contact angle
\begin{equation}
\vartheta = - \widetilde h_x(0,y) = - |k| \theta_e \zeta_k \cos ky .
\label{hth}  \end{equation}
Using this in (\ref{uth0}) yields the Fourier component of the contact line displacement velocity
\begin{equation}
 \dot\zeta_k =  - Q U_0 \theta_e^3 |k| \zeta_k.  
\label{zth}  \end{equation}
We see that the restoring force is proportional to $|k|$ rather than $k^2$ as it would be for a truly two-dimensional elastic line. This is in agreement with the qualitative argument \cite{deGennes85, JdeGennes84, Raphael01} stipulating that the capillary energy (ostensibly proportional to $k^2$) should be integrated over the penetration length of the order $|k|^{-1}$, thus leading to a reduced power of the wavenumber dependence. This makes the response of the contact line ``superdiffusive'', e.g. the curvature radius $r$ of a small bulge grows with time as $r \propto t^{2/3}$ rather than $r \propto t^{1/2}$. The relaxation rate proportional to $|k|$ is also supported by full numerical computations \cite{Snoeijer07}, which have shown that it breaks down only near the Landau--Levich entrainment limit. In this limit, the approximation breaks down, as the contact line perturbation computed as above becomes comparable with the unperturbed contact angle.

Another example to be studied in detail below is a circular droplet of the radius $R$. The unperturbed profile is a parabolic cap
\begin{equation}  
\label{hcap}
h_0(r) =\frac{\theta_eR}{2} \left[1 - \left(\frac{r}{R}\right)^2\right],
\end{equation}  
where $r$ is the radial coordinate $r$. The perturbation modes dependent on the polar angle $\phi$ are $\xi_n=\zeta_n \cos n\phi$ with an integer $n$. The solution of the Laplace equation for the perturbed interface is 
\begin{equation}
 \widetilde h(r,\phi) =  \theta_e\zeta_n \,(r/R)^n \cos n \phi ,
\label{hthr1}  \end{equation}
Taking also into account the contribution of $h_0'(r)$ yields the perturbation of the apparent contact angle at the contact line
\begin{equation}
\vartheta =  \theta_e  -h_0'(R+\xi_n) - \widetilde h_r(R,\phi)
 = - \frac{n-1}{R}\, \theta_e \zeta_n \cos n \phi .
\label{hthr}  \end{equation}
In a special case of a symmetric perturbation ($n=0$), the perturbation of $\theta$ is related to the perturbation of $R$ by the conservation condition of the droplet volume $V=\frac{4}{3}\theta_eR^3$, which yields $\vartheta =-\zeta_n\theta_e/R$. The expression (\ref{hthr}) can be extended also to this case if $n-1$ is replaced by its absolute value $|n-1|$.

\section{Dynamics of a diffusive surfactant} \label{S12}

A disturbance displacing the contact line may be caused by chemical inhomogeneities, which can be accounted for by assuming a linear dependence of the equilibrium contact angle on the local concentration of an adsorbed chemical:     
\begin{equation}
 \theta_e =  \theta_0 [1 - b (c - c_0)],
\label{cth}  \end{equation}
where $\theta_0$ is the reference contact angle at $c=c_0$ and $b$ is a proportionality constant.
Modification of wetting properties by adsorption or chemical reactions has been used in a number of experiments to induce spontaneous motion of droplets on solid substrate. In the following, we will consider a reversible setup allowing for restoration of substrate properties \cite{j05l}. Variation of the contact angle is caused in this system by deposition of a surfactant from a bulk phase (assumed to have negligible viscosity) and its dissolution underneath a droplet or liquid layer. The theoretical model describing stationary droplet motion in this setup \cite{p06} is based on a surface diffusion equation with a constant adsorption rate and linear desorption kinetics. This leads to the  surfactant diffusion equation written in the dimensionless form
\begin{equation}
c_t = \nabla^2 c - c +  H(\vect{x}) . 
\label{diff} \end{equation}
The surfactant coverage $c$ is scaled here by the coverage in equilibrium with the constant surfactant concentration in the continuous phase, time by the inverse desorption rate constant $k_d$, and length, by $\sqrt{D/k_d}$, where $D$ is the surfactant diffusivity on the substrate; $H(\vect{x})$ is the Heaviside step function equal to 1 outside and 0 inside the footprint of the droplet or liquid layer. For simplicity, it is assumed that the desorption rate is constant all over the substrate and the bulk concentration in the liquid layer is negligible.

We will apply now the same model in conjunction with Eq.~(\ref{uth}) to explore stability of a contact line. In this formulation, the problem is almost identical to that of stability of non-equilibrium structures in the FitzHugh--Nagumo or similar models \cite{book}. First, we have to find the stationary solution of Eq.~(\ref{diff}) for the unperturbed configuration. For a semi-infinte layer at $x<0$, the solution dependent on a single coordinate $x$ is
\begin{equation}
c_s(x) = \left\{ \begin{array}{lcc}
\frac{1}{2}\,\E^{x}    & \mbox{ at }  & x \le 0, \\
 1 - \frac{1}{2}\, \E^{-x}   & \mbox{ at }  &  x \ge  0.
\end{array} \right.
\label{3veq1s} \end{equation}
The reference concentration $c_0$ can be identified with the stationary concentration at the contact line $c_s(0)=1/2$. 

For a circular droplet, the stationary solution $c=c_s(r)$ of Eq.~(\ref{diff})  is
 \begin{equation}
c_s= \left\{ \begin{array}{lcc}
R K_1(R)I_0(r)   & \mbox{ at }  & r\le R \\
1 -  R I_1(R)K_0(r)]   & \mbox{ at }  & r\ge R, 
\end{array} \right.
\label{32soln} \end{equation}
where $I_n,\:K_n$ are modified Bessel functions. The values of $c_s(R)$ given by both lines of this formula are the same, in view of the identity $K_1(R)I_0(R) + K_0(R)I_1(R)= 1/R$. This common value should be identified with the reference concentration $c_0$. 

The concentration perturbation due to a contact line displacement can be computed in a simple way by observing that any shift of the contact line position by an increment $\xi \ll 1$ is equivalent to switching on or off the source term in Eq.~(\ref{diff}) in a narrow region near the contact line. This contributes to the equation for the perturbation $\widetilde c=c-c_s$ a source localized at the unperturbed contact line position and proportional to its displacement:
\begin{equation}
 \partial_t \widetilde c = \nabla^2 \widetilde c - \widetilde c - \xi \, \delta (\Gamma).
      \label{3sVeq}     \end{equation}

The dynamic equation for the contact line displacement is obtained by linearizing Eq.~(\ref{uth}). Using Eq.~(\ref{cth}) and expanding the surfactant concentration in the vicinity of the unperturbed contact line position yields $\theta_e =\theta_0[1 - b(\widetilde c + j \xi)]$, where $j$ is the stationary surfactant flux into the liquid across the contact line. Combining this with Eq.~(\ref{uth0})  yields the equation for the local contact line displacement  
\begin{equation}
 \dot\xi =  - \chi [\vartheta/\theta_0 - b(\widetilde c + j\,\xi) ].  
\label{zth1}  \end{equation}
We have transformed here to the dimensionless form based on the chemical length and time scales. The remaining dimensionless parameter $\chi =Q U_0 \theta_0^3/\sqrt{Dk_d}$ is proportional to ratio of the characteristic capillary velocity $U_0=\gamma/\eta$ to the characteristic chemical velocity $\sqrt{Dk_d}$. This ratio is typically very large, but is effectively reduced being multiplied by the cube of the contact angle, which must be small when the lubrication approximation is applicable. 

Although perturbations are habitually presumed to be arbitrarily small for the purpose of linear stability analysis, the method is actually also applicable to finite perturbations of the contact line position, provided they are small compared to the characteristic surfactant diffusion scale. 


\section{Instability of a rectilinear contact line} \label{S31}

For a rectilinear contact line displaced by a harmonic perturbation, $\vartheta$ in Eq.~(\ref{zth1}) is given by Eq.~(\ref{hth}) and the stationary flux following from Eq.~(\ref{3veq1s}) is $j= c_s'(0) = 1/2$. Presenting the concentration perturbation in the spectral form $\widetilde c(x,y,t) = \psi(x,t)\cos ky$ and setting in the perturbation equations (\ref{3sVeq}), (\ref{zth1}) $\psi, \zeta_k \sim \E^{\lambda t}$ leads to the eigenvalue problem 
\begin{eqnarray}
  \lambda \zeta_k &=&  - \chi [ |k| \zeta_k - b(\psi + \zeta_k/2) ] ,  
\label{zeq} \\
&& \psi_{xx} - q^2\psi = \zeta_k \delta (x),
      \label{3spsieq}     \end{eqnarray}
where $q^2=1 +\lambda + k^2$. The solution of Eq.~(\ref{3spsieq}), presuming Re $q>0$, is
\begin{equation}
\psi(x) = -\frac{\zeta_k}{2q} \, \E^{-q|x|}.
 \label{3sVsoln} \end{equation}
Using this in Eq.~(\ref{zeq}) yields the dispersion relation
\begin{equation}
  \lambda = - \chi \left[ |k|  - \frac{b}{2}  \left(1 -  \frac{1}{q}\right) \right] .
  \label{3disp0} \end{equation}
The stability condition is Re $\lambda<0$. As expected, $\lambda$ vanishes at $k=0$, which reflects the translational symmetry of the rectilinear line.  Unlike a similar problem in Ref.~\cite{book} with $k$ replaced by $k^2$, the derivative $\D \lambda/\D k$ at $k=0$ is always negative, so that the contact line is stable to long-scale perturbations.  Stability is lost as $\lambda$ vanishes at  
\begin{equation}
k = k_c = \sqrt{\frac{1+ \sqrt{5}}{2}}, \qquad b = b_c =\frac{(3+ \sqrt{5})^{3/2}}{ \sqrt{1+ \sqrt{5}}}.
 \label{bk} \end{equation}
A band of unstable modes around $k=k_c$ (which turns out to coincide with the golden ratio) widens at $b>b_c$. In dimensional units, the wavelength of growing perturbations is of the same order of magnitude as the diffusional length $\sqrt{D/k_d}$. 

\begin{figure}[t]
\begin{center}
\includegraphics[width=8cm]{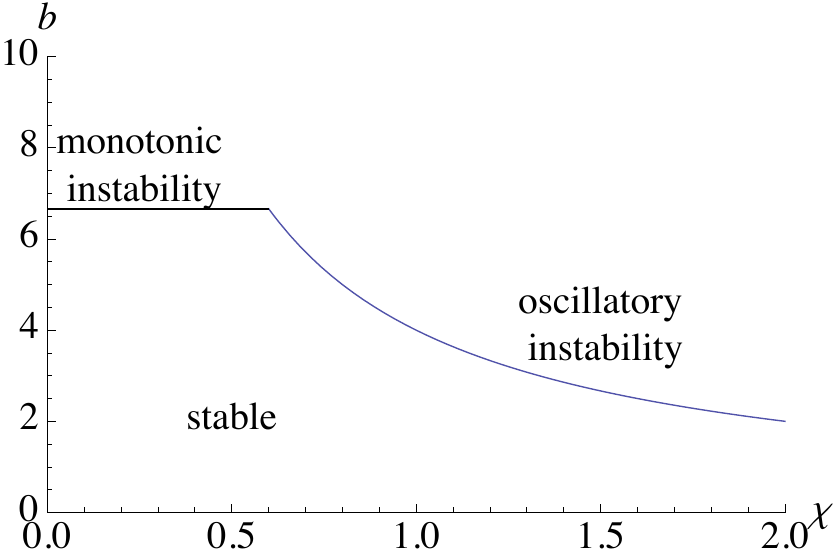} 
\end{center} \vspace{-5mm}
\caption{Monotonic and oscillatory instability regions in the parametric plane $\chi, b$ for a rectilinear contact line.}
\label{fstraight}
\end{figure}

In addition, oscillatory instability is possible when $\chi$ is sufficiently large. It first arises in the long-scale low-frequency mode at $\chi =\chi_c= 4/b$, and can be detected by expanding Eq.~(\ref{3disp0}) near this point. For  $\chi -\chi_c= \epsilon \ll 1$, the suitable scaling is 
\begin{equation}
\lambda = \epsilon  \lambda_r +\I \epsilon^{1/2} \widetilde \omega, 
 \qquad k = \epsilon \widetilde{k}.
 \label{keps} \end{equation}
The leading terms of the real and imaginary parts of the expansion are, respectively, of the order 1 and 3/2. This yields two lowest order equations for the real and imaginary parts of $\lambda$ as functions of $k$:
\begin{equation}
 \widetilde{k} - \frac{3}{16} b\, \widetilde \omega^2=0, \qquad
 b \, \widetilde \omega \left( 2b -12  \lambda_r -5\widetilde \omega^2 \right)=0.
  \label{keps1} \end{equation}
 The first of these gives the frequency as the function of the wavenumber:
\begin{equation}
 \omega= 4\sqrt{ \frac{k}{3b}}.
  \label{kom} \end{equation}
  The second equation gives the $O(\epsilon)$ real part of the eigenvalue
\begin{equation}
 \lambda_r = \frac{b}{6} -  \frac{20}{9} \frac{\widetilde k}{b},
  \label{klam} \end{equation}
indicating that instability indeed occurs at $\chi >\chi_c$ and disappears as $k$ increases. Higher orders of the expansion give $O(\epsilon^2)$ corrections to the real, and $O(\epsilon^{3/2})$ corrections to the imaginary parts of $\lambda$. Within the instability region, wave modes with finite wavelength also become unstable and are likely to exhibit there the fastest growth rate.  The monotonic and oscillatory instability regions in the parametric plane $\chi, b$ are shown in Fig.~\ref{fstraight}.

\section{Instability of a circular droplet} \label{S32}

For a circular droplet, $\vartheta$ in Eq.~(\ref{zth1}) is given by Eq.~(\ref{hthr}), and the eigenvalue problem for coupled spectral equations of the displacement and concentration perturbation analogous to Eqs.~(\ref{zeq}), (\ref{3spsieq}) has the common form for any integer $n \ge 0$:
\begin{eqnarray}
&&  \lambda \zeta_n =  - \chi [ |n-1| \zeta_n - b(\psi + j\,\zeta_n ) ] ,  
\label{zeqr} \\
&& \psi_{rr} + \frac{1}{r}\psi_r - \left(q^2 + \frac{n^2}{r^2}\right)\psi = 
        \zeta_n   \delta (r-R),
      \label{3spsieqr}     \end{eqnarray}
where $q^2=1 +\lambda$. The solution of Eq.~(\ref{3spsieqr}) is
\begin{equation}
\psi(r)= \left\{ \begin{array}{lcc}
- R \zeta_n  K_n(qR)I_n(qr)  &\mbox{ at } & r\le R ,\\
- R \zeta_n  I_n(qR)K_n(qr)  &\mbox{ at } & r\ge R. 
      \end{array} \right.
\label{3sVsolnd} \end{equation}
Using this, together with the expression for the stationary flux $j= c_s'(R) = R K_1(R)I_1(R)$ following from Eq.~(\ref{32soln}), in (\ref{zeqr}) yields the dispersion relation
\begin{equation}
\frac{\lambda}{\chi} = - \frac{|n-1|}{R} + b R \left[I_1(R)K_1(R) - I_n(qR)K_n(qR)\right] .
  \label{3disp0r} \end{equation}

\begin{figure}[t]
\begin{center}
\includegraphics[width=8cm]{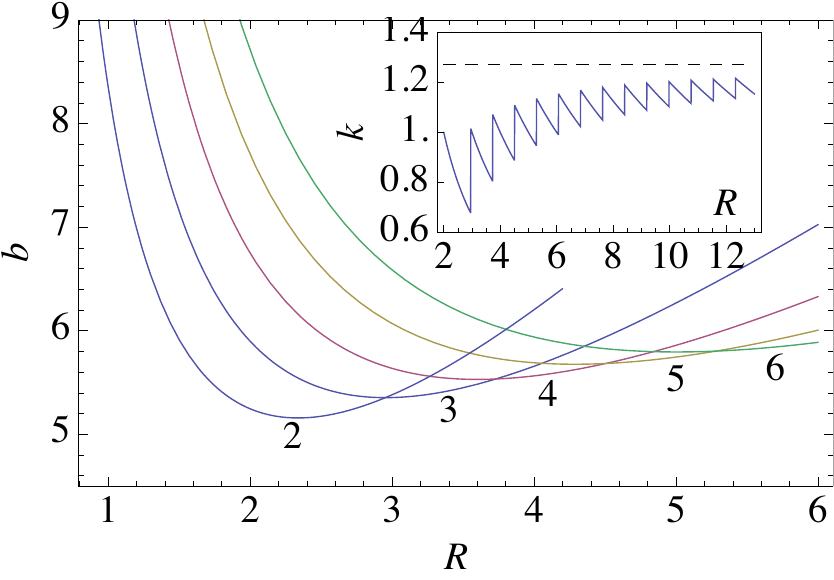} 
\end{center} \vspace{-7mm}
\caption{Monotonic instability thresholds for a circular droplet  of radius $R$. The numbers indicate the value of $n$. Instability occurs above the lower envelope of the family of curves.  Inset: the wavenumber $k=n/R$ for the most dangerous mode; the dashed line indicates the respective value for the rectilinear contact line.}
\label{fcircle}
\end{figure}

As expected, $\lambda$ vanishes at $n=1$, which corresponds to a shift without deformation. For $n \neq 1$,  the critical value $b=b_n$ for a given radius verifies Eq.~(\ref{3disp0r}) with $\lambda=0$:
\begin{equation}
\frac{1}{b_n} = \frac{R^{2}}{|n-1|} \, \left[I_1(R)K_1(R) -I_n(qR)K_n(qR)\right] .  
 \label{posbc} \end{equation}
For $n =0$, the value $b_0$ given by this relation is negative, so that no monotonic instability can arise in the symmetric mode. The most dangerous instability mode is the one with the lowest critical value of $b_\perp(R)=\min_n b_n(R)$. For small droplets, the dipole mode deforming the circular droplet into an ellipse, is the most dangerous one. With the growing radius, the most dangerous monotonic instability mode shifts to larger values of $n$, as shown in Fig.~\ref{fcircle}. The absolute lower limit of monotonic instability is $b_2 \approx 5.158$ at $R\approx 2.337$, which is below the respective limit $b_c \approx 6.6604$ for the rectilinear contact line given by Eq.~(\ref{bk}); the latter limit is approached asymptotically at $R \to \infty$, as the envelope of the family of curves in Fig.~\ref{fcircle} gradually rises. The wavenumber $k=n/R$ rises as well on the average, as seen in the inset in the same Figure, coming gradually closer to the value $k_c \approx 1.272$ of Eq.~(\ref{bk}). 

\begin{figure}
\begin{center}
\includegraphics[width=8cm]{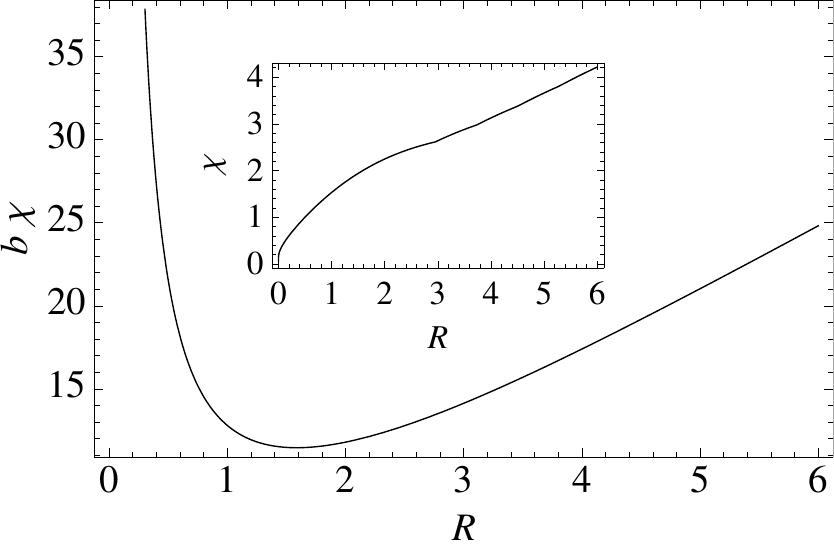} 
\end{center} \vspace{-7mm}
\caption{The lower limit of the product $b\chi$ at the traveling instability limit as a function of radius $R$. Inset: The lower limit $\chi_t$ at which the traveling instability precedes the monotonic one. }
\label{fmove}
\end{figure}

Besides the monotonic instability, dynamic instabilities dependent on the parameter $\chi$ are possible. Those are traveling instability setting the droplet into motion and oscillatory or wave instability. 
The threshold of traveling instability is detected, according to the general algorithm \cite{book}, by differentiating the dispersion relation (\ref{3disp0r}) for $n=1$ with respect to $\lambda$ and solving the resulting equation at $\lambda=0$. The droplet sets into spontaneous motion at 
\begin{equation}
  R[I_0(R)K_1(R)- I_1(R)K_2(R)]> 2/b \chi.
\label{3ct2} \end{equation}
The lowest value of the product $b \chi$ enabling traveling instability $b \chi \approx 11.46$ is attained at $R \approx 1.5866$ (see Fig.~\ref{fmove}). The lower limit $\chi=\chi_t$ at which the traveling instability precedes the monotonic one for droplets with a certain radius can be obtained by using in Eq.~(\ref{3ct2}) the critical value $b_\perp$ for the most dangerous mode given by Eq.~(\ref{posbc}). The plot $\chi_t(R)$ is shown in the inset of Fig.~\ref{fmove}.

Oscillatory instability can be detected numerically by solving coupled equations for the real and imaginary parts of Eq.~(\ref{3disp0r}). Setting $\lambda = \I\omega$ we obtain at the instability threshold
\begin{eqnarray}
&& \frac{1}{b} = \frac{R^{2}}{|n-1|} \, \left[I_1(R)K_1(R) -\mbox{Re}\,F_n(R,\omega)\right],  \label{posb} \\
&&\frac{1}{\chi} = -\frac{bR}{\omega} \,\mbox{Im} \, F_n(R,\omega),
   \label{posc} \\
&& F_n(R,\omega) = 
   I_n \left(R\sqrt{1+\I\omega} \right)
              K_n \left( R\sqrt{1+\I \omega}\right).
  \label{posf} \end{eqnarray}
Using Eq.~(\ref{posb}) to eliminate $b$, we are left with a single equation (\ref{posc}), which can be solved numerically to compute the frequency $\omega$ at the Hopf bifurcation point for a droplet of a given radius $R$. 

For $n \ge 2$, the locus of oscillatory instability branches off the locus of monotonic instability defined by  Eq.~(\ref{posbc}) at the point of double zero eigenvalue where the frequency vanishes. Close to this line, $\omega$ is small, and a Taylor series can be used. The leading term in the expansion of Eq.~(\ref{posb}) is of the order $O(\omega^2)$, while the leading term in the expansion of Eq.~(\ref{posc}) is linear in $\omega$. Setting the latter to zero gives the value of $\chi=\chi_n$ required to obtain the double zero as a function of the radius $R$; the same relation can be obtained most readily by differentiating Eq.~(\ref{3disp0r}) with respect to $\lambda$. The resulting condition is a generalization of  Eq.~(\ref{3ct2})
\begin{equation}
 R[I_{n-1}(R)K_n(R)- I_n(R)K_{n+1}(R)]  > 2/b \chi_n . 
\label{3ct2d} \end{equation}
The value of $\chi_n$ given by this relation consistently rises with $n$, and is always higher than the respective value for traveling instability obtained for $n=1$. This proves that low-frequency oscillatory instability never occurs before traveling instability sets in.

The remaining possibility is oscillatory instability with a finite frequency, which is also feasible (and, indeed, is most likely to occur) in a symmetric mode. Rather than solving Eqs.~(\ref{posb}), (\ref{posc}) directly, it is more instructive again to compute the lower limit $\chi=\chi_o$ at which this instability  precedes the monotonic one for a droplet with a certain radius. For this purpose, $\omega$ is computed by solving numerically Eq.~(\ref{posb}) with $b=b_\perp$, and then $\chi_o$ is evaluated using Eq.~(\ref{posc}). The computation shows that the frequency at the Hopf bifurcation decreases monotonically, going down to zero as the limit of a rectilinear contact line approaches, in agreement with the results of Sect.~\ref{S31}. The curve $\chi_o(R)$ for the symmetric mode lies, however, slightly higher than the curve $\chi_t(R)$, so that traveling instability always occurs first.

\section{Instability of a moving contact line} \label{S4}

The analysis can be extended to the case when the difference between the equilibrium and apparent  contact angles is not small and the unperturbed line is moving with a certain velocity $U$. The deviation $\vartheta$ is now redefined as a perturbation of the apparent contact angle at the moving contact line $\theta$, which, in accordance with Eq.~(\ref{uth}), equals to $( \theta_e^3 + 3U/QU_0)^{1/3}$. In the same approximation as in Sect.~\ref{S11}, $\vartheta$ is given for a rectilinear contact line by Eq.~(\ref{hth}) with $\theta_e$ replaced by $\theta$. The equation of  the local contact line displacement (\ref{zth1}) has to be written now for the displacement $\xi$ relative to the unperturbed moving line, and is modified to
\begin{equation}
 \dot\xi =  -  \left(\chi+3u \right) \,\vartheta/\theta
   + b\chi(\widetilde c + j\,\xi) ,
\label{zthv}  \end{equation}
where $u=U/\sqrt{k_dD}$ is the dimensionless velocity. 

The flux through the contact line defined by solving Eq.~(\ref{diff}) in the comoving frame is $j= c_s'(0) = 1/\sqrt{4+u^2}$, and the spectral equations (\ref{zeq}), (\ref{3spsieq}) are modified to
\begin{eqnarray}
  \lambda \zeta_k &=&  -\left( \chi+ 3u \right) |k| \, \zeta_k 
  + b\chi\left(\psi + \frac{\zeta_k}{\sqrt{4+u^2}}\right) ,  
\label{zeqv} \\
&& \psi_{xx} - u\psi_x - q^2\psi = \zeta_k \delta (x).
      \label{3spsiv}     \end{eqnarray}
The solution of the last equation is
\begin{equation}
\psi(x) = -\frac{\zeta_k}{\sqrt{4q^2+u^2}} \,
 \exp \left[-\frac{x}{2} \left(u \pm \sqrt{4q^2+u^2 }\right) \right] , 
 \label{3sVsolnv} \end{equation}
where the positive and negative signs apply, respectively, at $x>0$ and $x<0$. Using this in Eq.~(\ref{zeqv}) yields the dispersion relation
\begin{equation}
  \lambda = - \left( \chi+ 3u \right)|k| 
   +  b\chi  \left( \frac{1}{\sqrt{4+u^2}}
     -  \frac{1}{\sqrt{4q^2+u^2 }}\right)  .
  \label{3dispv} \end{equation}
Notably, the coefficient at $|k|$ vanishes at  $u=-\chi/3$, which corresponds to the Landau--Levich entrainment limit; the theory becomes inapplicable close to this point \cite{Snoeijer07}.  Besides the coefficient at $|k|$, the dispersion relation is insensitive to reversing the direction of motion, but the reference concentration $c_0$ defined as the surfactant concentration on the unperturbed contact line, is different in the two cases. 

\begin{figure}[t]
\begin{center}
\includegraphics[width=8cm]{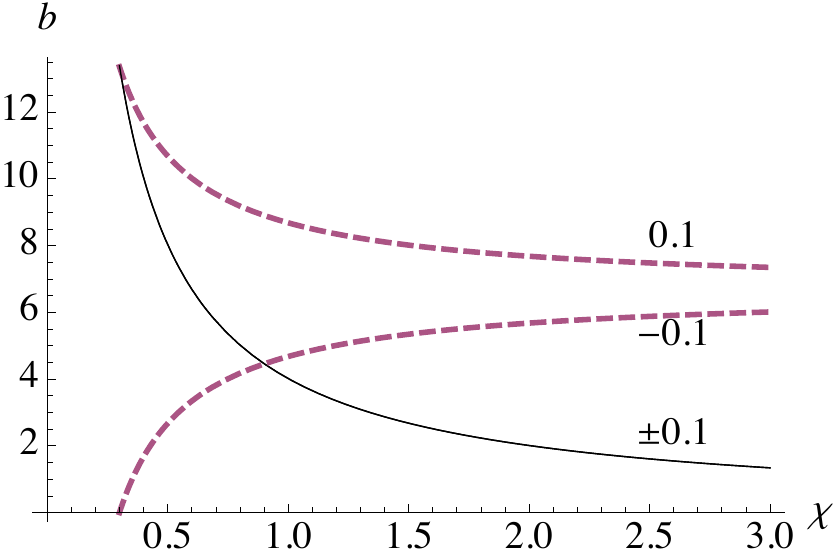} 
\end{center} \vspace{-7mm}
\caption{Loci of monotonic (dashed lines) and oscillatory (solid line) instability in the parametric plane $\chi, b$ for the rectilinear contact line moving with the speed $u=\pm0.1$ (as indicated by numbers at the curves). The stable domain is in below both respective curves.}
\label{fv}
\end{figure}

Monotonic instability is observed at  
\begin{eqnarray}
 k_c &=& \sqrt{\frac{1}{2}\left( 1+ \sqrt{5}\right) \left(1+ \frac{u^2}{4} \right)}, \label{kv} \\
  b_c &=& \frac{\left(3+\sqrt{5}\right)^{3/2} \left(1+u^2/4\right) (3 u+\chi)}{\chi
   \sqrt{1+\sqrt{5}} }.
 \label{bv} \end{eqnarray}
Besides the velocity dependence, a qualitatively significant change, compared to  Eq.~(\ref{bk}), is the dependence of the monotonic instability threshold on the parameter $\chi$ which appears at any non-zero velocity; the marginal wavelength $k_c$ remains, however, independent of $\chi$. At $\chi \to 0$, the velocity dependence becomes singular: a retreating contact line $u<0$  is always unstable under these conditions, while for an advancing line ($u>0$) the threshold rises sharply.

Oscillatory instability emerges, as at $u=0$, in a long-wave low-frequency mode, and can be detected by expanding Eq.~(\ref{3dispv}) using the scaling (\ref{keps}) as before. The $O(\epsilon^{1/2})$ term gives the parametric relation at the instability threshold 
\begin{equation}
b\chi= 4(1+u^2/4)^{3/2} . 
 \label{bchiv} \end{equation}
The dispersion relation $\omega( k)$ following from the $O(\epsilon)$ term is
\begin{equation}
 \omega=\sqrt{k \left(u^2+4\right) \left[ b+\frac{\left(u^2+4\right)^{3/2}}{6u}\right]}.
  \label{komv} \end{equation}
For $u>0$, this relation is qualitatively similar to Eq.~(\ref{kom}); the function $\lambda_r (\widetilde k)$ defined by the $O(\epsilon^{3/2})$ term is likewise similar to Eq.~(\ref{klam}). The overall bifurcation diagram changes, however significantly, compared to that in Fig.~\ref{fstraight}. The loci of monotonic and oscillatory bifurcation fail to intersect already at $u > 0.2012$; at higher velocities, only oscillatory  instability is relevant.  At $u < 0$, on the contrary, monotonic instability prevails at small $\chi$ (see Fig.~\ref{fv}). 

\section{Discussion} \label{S5}

We have seen in Sect.~\ref{S1} a remarkable similarity between the contact line equilibrium (described in lubrication approximation) and equilibrium of an interphase boundary in two dimensions. Likewise, chemo-capillary instabilities of the contact line driven out of equilibrium by coupling with surfactant adsorption (or other non-equilibrium process affecting wetting properties of the substrate) are remarkably similar to instabilities of two-dimensional non-equlibrium structures. A substantial difference stems, however, from relaxation response of a curved contact line being proportional to the absolute value rather than square of the wavenumber $k$. This may be attributed to a ``two-and-a-half-dimensional'' character of the lubrication approximation. In view of this specific response, the concept of line tension lifted from the two-dimensional world is not duly applicable to the contact line. This is manifested, in particular, in the paradox of stability of the contact line at an apparently negative line tension \cite{Dietrich07,Mechkov07}.    

Out of equilibrium, the modified wavenumber dependence brings about more subtle changes, compared to instabilities of dissipative structures in two dimensions \cite{book}. The onset of instability of a rectilinear contact line shifts to a finite wavelength, and traveling instability prevails over the oscillatory one for circular droplets. We have seen in Sects.~\ref{S31}, \ref{S32} that monotonic instability occurs on the wavelength of the same order of magnitude as the surfactant diffusion length. It should cause fingering in the case of macroscopic liquid volumes of a size far exceeding this length, like it is commonly observed in spreading of surfactant-laden liquid layers \cite{matar}. It can also explain blebbing and fission of moving droplets observed in the experiment \cite{Nagai}. Instabilities with lower $n$ may result in splitting of microdroplets with a radius of the same order of magnitude as the diffusion length. Dynamic instabilities are enhanced by the increased ratio of the characteristic hydrodynamic and reaction-diffusion velocities expressed by the parameter $\chi$. One should be aware, however, that the variety of instabilities predicted by theory may be masked in actual experiments by surface roughness arresting contact line displacement.


\end{document}